\documentclass{article}

\usepackage{arxiv}

\usepackage{algorithm}
\usepackage{algorithmic}
\usepackage{graphicx}
\usepackage{amsthm}
\usepackage{indentfirst}
\usepackage{algorithm}
\usepackage{algorithmic}
\usepackage{multirow}
\usepackage[toc,page]{appendix}
\usepackage{algorithm,float}
\usepackage{amssymb}
\usepackage{makecell}
\usepackage{array}
\usepackage{mathrsfs}
\usepackage{amsmath}
\usepackage{amssymb}
\usepackage{mathrsfs}
\usepackage{CJK}
\usepackage{multicol} 
\usepackage{hyperref}

\def\cA{{\cal A}}
\def\cC{{\cal C}}
\def\cG{{\cal G}}
\def\cK{{\cal K}}
\def\bbZp{\mathbb{Z}_{p}}

\usepackage{amsmath}

\newtheorem{definition}{Definition}

\makeatletter

\newcommand{\beit}{\begin{itemize}}
	\newcommand{\eit}{\end{itemize}}

\title{Provenance-based Classification Policy based on Encrypted Search}

\author{Xinyu Fan\thanks{Corresponding author}, Faen Zhang, Jiahong Wu, Jingming Guo\\
fanxinyu@ainnovation.com zhangfaenainnovation@gmail.com\\
\{wujiahongainnovation, guojingmingainnovation\}@gmail.com\\
AInnovation Technology Ltd.} 

\begin{document}
\maketitle

\begin{abstract}
	As an important type of cloud data, digital provenance is arousing increasing attention on improving system performance. Currently, provenance has been employed to provide cues regarding access control and to estimate data quality. However, provenance itself might also be sensitive information. Therefore, provenance might be encrypted and stored in the Cloud. In this paper, we provide a mechanism to classify cloud documents by searching specific keywords from their encrypted provenance, and we prove our scheme achieves semantic security. 
	In term of application of the proposed techniques, considering that files are classified to store separately in the cloud, in order to facilitate the regulation and security protection for the files, the classification policies can use provenance as conditions to determine the category of a document. Such as the easiest sample policy goes like: the documents have been reviewed twice can be classified as "public accessible", which can be accessed by the public.
\end{abstract}
\section{Introduction}
The definition of provenance in the Oxford English Dictionary is: \emph{(1) the fact of coming from the particular source or quarter; origin, derivation; (2) the history or pedigree of a work of art, manuscript, rare book, etc.; concretely, a record of the ultimate derivation and passage of an item through its various owners.} In computer systems, the provenance of data refers to the logs of the processes and operations of the data, which is relevant to its sources and origins. Provenance can be expressed as a directed acyclic graph (DAG),  illustrating how a data artifact is processed by an execution. In such a DAG of provenance under the Open Provenance Model (OPM)\cite{MoreauCFFGGKMMMPSSB11}, nodes present three main entities including \emph{Artifact}, \emph{Agent} and \emph{Process} and edges represent connections to the main entities.

Data provenance records historical operations performed on documents, preserving its security and privacy. Provenance Access Control is considered a crucial research topic for big data security. The sensitivity of files and their provenance can be different, and users can request, and be granted, access to files and provenance separately. In some situations, provenance itself may consist of sensitive information which might require more protection than its attached document. For instance, although a programming project can be published to the public, its authors and executed operations should be kept as a secret, to prevent leaking the techniques. Therefore, access control to the provenance data itself is required. It allows eligible users to access the provenance data and protects it from unauthorised access. 

A number of proposals for access control based on the provenance of the data have been made in some papers\cite{NguyenPS13}\cite{ParkNS12}. In terms of access control, characteristics such as data accuracy, timeliness and the path of transfer of data can be crucial restrictions of a policy. With increasing regulation, the consequences for signing incorrect statements have significantly increased, even if the signer was not directly responsible for the invalid sections. Therefore it is important to track which entities were responsible for the process that led to the final form of the data.

The application of the technique proposed in this paper is that data are classified based on its provenance, in order to facilitate the data management and maintain. Admittedly, processes performed on data can provide clues to identify the vulnerability of data. In this system, we assume that date was performed under same or similar operations should be classified as a category. The access control should be performed based on the classification. For example, if data was collected by policemen for the purpose of detection, the collected data can be identified as very sensitive data. It was labeled as ``sensitive class", and its privacy should be protected carefully. The possible protection methods might include generating restricted access policies and encrypt it.

\section{Our Contributions}

In this paper, we consider the approach where the data carries with it provenance information which can be used to make access control decisions. A generic representation is sufficiently extensible to capture the essence of the semantics of provenance across various domains. If such a representation can be captured in a secure manner, then it will be useful in tackling the issue of attribution of data as it moves around the cloud. For instance, information about the origin of data together with the conditions and the state under which it was created along with the modifications that have been made and the conditions under which these modifications have been made will allow the access control service to more robustly make security decisions. Such an approach would transform the access control service to a more stateful decision and make it more context-dependent.

In cloud storage systems, to provide appropriate management and security protection, files could be stored in separate units. While each unit keeps a category of files, classifying files as categories is an effective mechanism for organising files and management access to files. In this paper, we focus on classifying files according to provenance which records generating a process of files. Specifically, our system identifies and classifies files by their own preferences based on which process worked on which files. For example, when medical records or governmental survey documents are anonymised, it removes sensitive personal information. Then, these files could be accessed by the public, students, and scholars for the purpose of research. Therefore, it identifies files after anonymization as ``public education". On the contrary, if files are combined with judgment or comments with sensitive agents, they might wish to keep these files secure from access to with the public or attackers. Then these files are classified as ``sensitive information" and take a higher level protection and deny access from unauthorised users. However, the third party to execute classification might not be fully trustworthy either. To prevent an internal attack and retain confidentiality of data, we hope to keep data information confidential by encryption as well as classification them. We provide a provenance-based classification system to implement this goal. In this system, we propose a scheme to search keywords from encrypted provenance. When specific keywords are found, files are classified by according system policies. 

In particular, our paper makes the following contributions: 

\begin{itemize}
	
	\item Allowing the policy decision server to check the encrypted provenance without decrypting the provenance, while at the same time;
	
	\item Providing guarantees to the policy decision server that the provenance is from a genuine source and is linked with the particular data or file. 
	
\end{itemize}  

Such a solution will enable authenticated and confidential provenance information to be used in the access control service without revealing its plain content. To achieve such a solution, we introduce a new notion of Encrypted Provenance Search Scheme (EPSS). EPSS is based on the searchable encryption method proposed by Boneh \emph{et al.}\cite{BonehF01}. \\

\subsection{Paper Organization}\label{sesddsd}
The organisation of this paper is as follows.

\begin{itemize}
	\item Section 2 briefly presents some research works in the areas of provenance and encrypted search that are relevant to our work. 
	
	\item Section 3 gives a brief introduction to the representation of provenance and its characteristics. 
	
	\item Section 4 presents our Provenance-based Classification Access Policy (PBCAP) and system architecture, which is followed by preliminaries in section 5. 
	
	\item Section 6 proposes the Provenance-based Classification Scheme
	
	\item Section 7 presents semantic-secure game for it and proves that this scheme is semantic secure. 
	
	\item Finally, Section 8 concludes the paper and states the anticipated future work. 
	
\end{itemize}

\subsection{Related Work}
Data provenance might be around sensitive information and in those cases, the security of provenance (for example \cite{BraunSS08}, \cite{BertinoGKNPSSTX14} and \cite{HussainWSB14}) has aroused increasing attention. There are several attempts to encrypt provenance information to keep its confidentiality. Li \emph{et al.}\cite{LiCHW14} proposed a provenance-aware system based on Attribute-based signature (ABS) which supports fine-grained access control policies. The users' privacy is also protected because attribute private key of users is issued with an anonymous key-issuing protocol from multiple attribute authorities. However, the whole computation is built on the assumption that the server has a large computational ability. Chow \emph{et al.} \cite{ChowCHZD12} propose a cryptographic design for cloud storage systems supporting dynamic users and provenance data. These encryption schemes could contribute the system we proposed in this paper, however, our contribution focus on serving the classification policies.

In the area of encrypted data search, Boneh \cite{BonehCOP03} presents a Public Key Encryption with Keyword Search (PKES) scheme. We will be making use of this work in the design of our Provenance-based Classification Access scheme. Essentially, the work in paper \cite{BonehCOP03} considers the following scenario: when Alice receives emails, she would like to set a gate that helps her to check whether the incoming emails contains certain sensitive keywords such as ``urgent". However, the emails are encrypted to protect privacy. As the gateway is not fully trusted, Alice does not want to grant the gateway the ability to decrypt her emails.  The PKES scheme enables the gateway to conduct a test to verify if the encrypted emails contain the keywords while learning nothing else about the content of the emails themselves.

\section{System Architecture and the Policies}

\subsection{System Architecture}
We assume that a cloud service provider has several remote storage units available for the storing data that is received from different users of the system. Users wish to store their files with provenance in the cloud and send them in an encrypted format to the cloud service provider. The goal is to design an access control system that the cloud service provider can use to classify encrypted files by searching keywords from encrypted provenance. We refer to this access control system as a Provenance-based Classification Access Policy (PBCAP) System. 

Figure 1 gives an outline of our system architecture. The remote data storage units are managed by a cloud server which classifies files and allocates them in corresponding storage units.  Our PBCAP system achieves the following objectives: (1) the cloud server will classify the encrypted files that it receives from users based on the attached encrypted provenance information; (2) the encrypted provenance information is checked for policy compliance while they remain encrypted (hence the confidentiality of both the encrypted files and their provenance information are guaranteed); and (3) provides a guarantee to the cloud server that the provenance is from a genuine source. \\

\begin{figure*}[thb]
	\vspace{-0cm}
	\centering
	\includegraphics[scale=0.3]{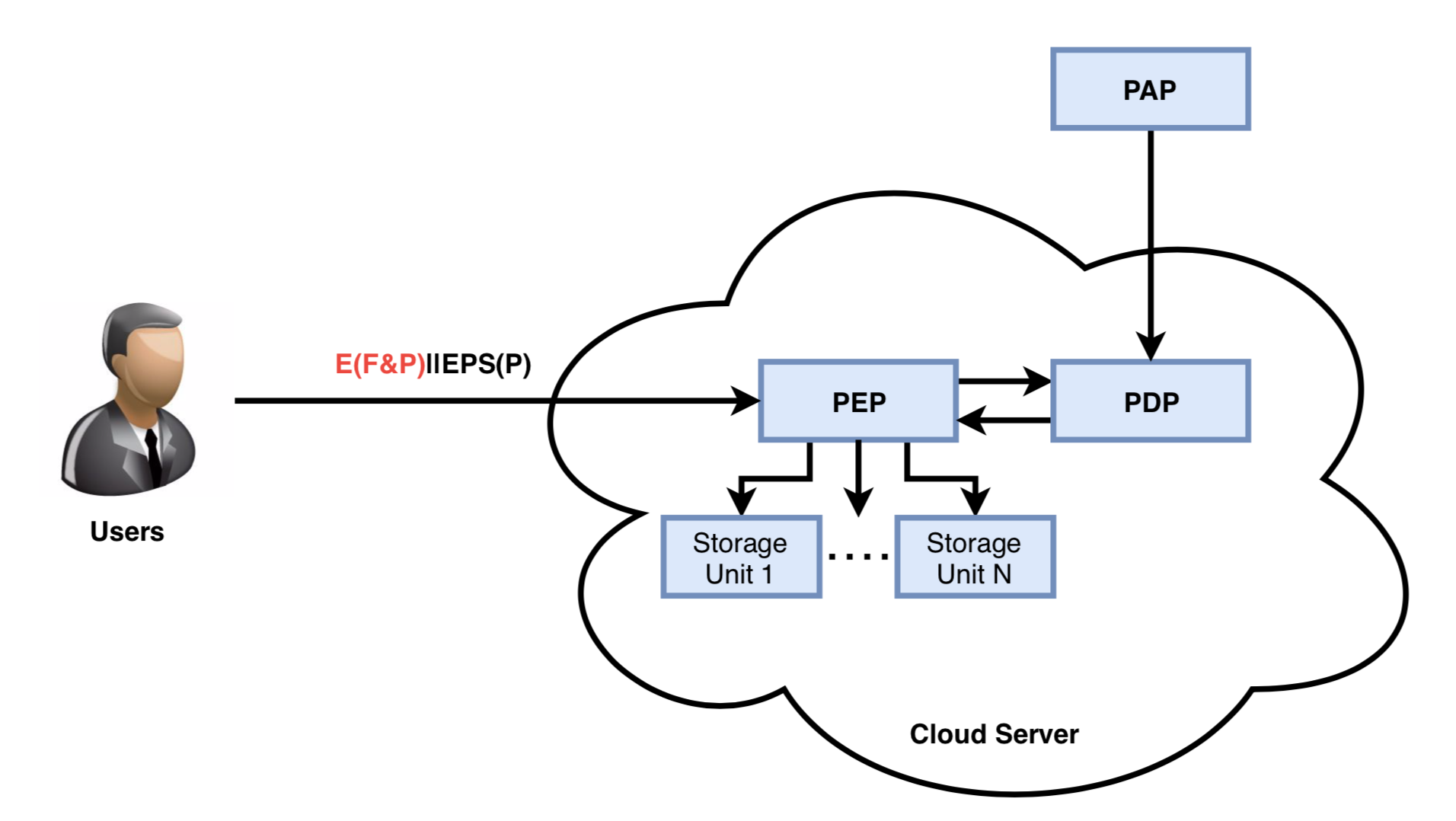}
	\vspace{-0.2cm}
	\caption{PBCAP System Architecture} \label{fig1}
\end{figure*}

The components of the system architecture are as follows:

\begin {itemize}
\item {\sf Users} are the owners of files who send encrypted files to the cloud server for storage. The files along with their provenance are encrypted by the user before they are sent to the cloud server.  In our scheme, users also generate a pair of public and private keys which are used for an embedded short signature verification mechanism. \\
\item {\sf Cloud Server} classifies the received encrypted files and stores them in different storage units. Each unit stores a number of files that share common attributes. For instance, an attribute might be undergoing a specific process (e.g. being graded by Alice, or being edited by Bob). This benefits management of files as well as providing corresponding levels of security protections. \\
\item {\sf Policy Administrator Point (PAP)} generates policies and sends them to Policy Decision Points (PDP) for implementation. To keep confidentiality, PAP encrypts sensitive information in policies before sending them to PDP.  \\
\item {\sf PDP} Executes Encrypted Provenance Search Scheme (EPSS) as per the following the steps: verifying short signatures of provenance ciphertexts to make sure they are from genuine users; searching keywords from encrypted provenance and output results to Policy Enforcement Point (PEP). \\
\item {\sf PEP} receives results from PDP and allocates files to corresponding storages units. \\
\end {itemize}

\subsection{Provenance-based Classification Policy}

Provenance-based classification policies classify files based on their provenance. Namely, each policy defines that files executed by a given of operations can be classified as one category. Hence, a policy maps a set of provenance partitions to a category. In terms of the motivation, operations performed on data can reveal its sensitivity and vulnerability. Hence, we employ provenance partitions as keywords to classify data.

To avoid conflict, we define different priorities for the classification policies. We give an example policy as below showing that if the provenance of a document includes any provenance partitions in the given set including RecordedBy (Test, Nurse), DiagnosedBy (Report, Doctor), the document will be classified as a medical document. Further, it will be stored in Hospital Storage unit which can only be accessed by staff and patients from the hospital. Particularly, the priority is $P_i$ of this policy, which implies that when a provenance consists of keywords in policy ID=1 and other policies, the data connected with the provenance should be classified as a category of policy with the highest priority among them. \\  

\begin{algorithm}
	\caption{Policy }
	\label{alg:A}
	\begin{algorithmic}
		\STATE {$<$Policy ID = $``1">$} 
		\STATE {$<$Typed Provenance Set$>$ RecordedBy (Test, Nurse), \\ DiagnosedBy (Report, Doctor)} 
		\STATE {$<$/Typed Provenance Set$>$} 
		\STATE {$<$Priority$>$ $P_i$ $<$/Priority$>$} 
		\STATE {$<$Category$>``$Medical Documents"  $<$/Category$>$} 
		\STATE {$<$Cloud Storage Unit$>``$Hospital"$<$/Cloud Unit$>$\\
			$<$/Policy$>$} 
	\end{algorithmic}
\end{algorithm}

\subsection{Public-Key Encryption}\index{Public-Key Encryption}\index{PKE}
Diffie and Hellman \cite{dh:76} introduced new research directions in cryptography called {\em public-key cryptography} (PKC)
where two parties can communicate over public channels without compromising the security of the system.

A public-key (asymmetric) encryption (PKE) scheme is a public-key cryptographic scheme used to protect the confidentiality of the transferred messages. In a PKE scheme, a secret public key pair is generated. Notably, it is computationally infeasible to obtain the secret key from the public key. This is in contrast with a symmetric encryption scheme where both the { decryption} key and the encryption key are same or it is easy to compute one from the other.

The formal definition of a PKE scheme is as follows \cite{dh:76}.  A PKE scheme consists of the following four algorithms.
\begin{itemize}
	\item{\sf Setup} $(1^{\ell})\rightarrow params.$ The setup algorithm takes as input $1^{\ell}$ and outputs the public parameters $params$.
	\item{\sf KeyGen} $(1^{\ell})\rightarrow(SK,PK).$ The key generation algorithm takes as input $1^{\ell}$ and outputs a secret-public pair $\cK\cG(1^{\ell})\rightarrow(SK,PK)$.
	\item{\sf Enc} $(ps,PK,M)\rightarrow CT.$ The encryption algorithm takes as input the public parameters $params$,  the public key $PK$ and a message $M$, and outputs a ciphertext $CT$.
	\item{\sf Dec} $(params, SK, CT)\rightarrow M.$ The decryption  algorithm takes as input the public parameters $params$,  the secret key $SK$ and the ciphertext $CT$, and outputs the message $M$.
\end{itemize}

\begin{definition}{\sf Correctness.}
	We say that a public-key encryption scheme is correct if
	\begin{eqnarray*}
		\Pr\left[\begin{array}{l|l} & {\sf Setup}(1^{\ell})\rightarrow ps;\\
			{\sf Dec}(ps,SK,CT)\rightarrow M & {\sf KeyGen}(1^{\ell})\rightarrow(SK,PK);\\
			& {\sf Enc}(ps,PK,M)\rightarrow CT\end{array}\right]=1
	\end{eqnarray*}
	where the probability is taken over the random coins consumed by all algorithms in the scheme.
\end{definition}

\noindent{\em Security Model.} The standard notion of the security for a PKE scheme is called {\em indistinguishability against adaptively chosen ciphertext attacks} (IND-CCA2)\cite{rs:91}.
This model is defined by the following game executed between a challenger $\cC$ and an adversary $\cA$.
\begin{itemize}
	\item{\sf Setup.} $\cC$ runs  {\sf Setup} $(1^{\ell})$ to generate the public parameters $params$ and sends them to $\cA$.
	\item{\sf KeyGen.} $\cC$ runs  {\sf KeyGen} $(1^{\ell})$ to generate the secret-public key pair $(SK,PK)$ and sends the public key $PK$ to $\cA$.
	\item{\sf Phase 1.} $\cA$ can adaptively query the decryption oracle. $\cA$ submits a ciphertext $CT$ to $\cC$, where $CT=Enc (param,PK,M)$. $\cC$ runs {\sf Dec} $(params,SK,CT)$  and responds $\cA$ with $M$. This query can be made multiple times.
	\item{\sf Challenger.} $\cA$ submits two messages $M_{0}$ and $M_{1}$ with equal length. $\cC$ randomly selects $M_{b}$  and computes $CT^{*}={\sf Enc}(params,PK,M_{b})$, where $b\in\{0,1\}$. $\cC$ responds $\cA$ with $CT^{*}$.
	\item{\sf Phase 2.} $\cA$ can adaptively query the decryption oracle. $\cA$ submits a ciphertext $CT$ to $\cC$, where the only restrict is $CT\neq CT^{*}$. {\sf Phase 1} is repeated. This query can be made multiple times.
	\item{\sf Guess.} $\cA$ outputs his guess $b'$ on $b$. $\cA$ wins the game if $b'=b$.
\end{itemize}

\begin{definition} {\sf IND-CCA2.}\index{Indistinguishability against Adaptive \par Chosen Ciphertext Attacks}\index{IND-CCA2}
	We say that a public-key encryption scheme is $(T,q,$ $\epsilon(\ell))$-indistinguishable against adaptive chosen ciphertext attacks (IND-CCA2)  if no PPT adversary $\cA$ making $q$ decryption queries can win the game with the advantage
	\begin{eqnarray*}
		Adv_{\cA}^{IND-CCA2}=\left|\Pr[b'=b]-\frac{1}{2}\right|\geq \epsilon(\ell)
	\end{eqnarray*}
	in the above model.
\end{definition}

Another security notion for public-key encryption is called {\em indistinguishability against adaptively chosen plaintext attacks} (IND-CPA). In this model, the adversary $\cA$ is not allowed to query the decryption oracle. The formal definition of this model is as follows.

\begin{definition}{\sf IND-CPA.}\index{Indistinguishability against  Adaptive \par Chosen Plaintex Attacks}\index{IND-CPA}
	We say that a public-key encryption scheme is $(T,\epsilon(\ell))$-indistinguishable against  adaptive chosen plaintex attacks (IND-CPA)  if no PPT adversary $\cA$ who is restricted to query the decryption oracle  can win the game with the advantage
	\begin{eqnarray*}
		Adv_{\cA}^{IND-CPA}=\left|\Pr[b'=b]-\frac{1}{2}\right|\geq \epsilon(\ell)
	\end{eqnarray*}
	in the above model.
\end{definition}

Some well known PKE schemes include the ElGamal encryption scheme
\cite{elg:85}, RSA encryption scheme \cite{rsa:78}, Cramer-Shoup
encryption scheme \cite{cs:98} and  RSA-OAEP encryption scheme
\cite{fops:01}.

\subsection{Digital Signature}\index{Digital Signature}
Digital signature was proposed by Diffie and Hellman \cite{dh:76}. It is the {electronic} version of a handwritten signature. A valid digital signature can convince a verifier that it was generated by a known party for a public message. Especially, a digital signature can provide non-repudiation property, namely, a signer cannot deny he has generated the signature.

A digital signature scheme is formally defined as follows \cite{gmr:88}. It consists of the following four algorithms.
\begin{itemize}
	\item{\sf Setup} $(1^{\ell})\rightarrow params.$ The setup algorithm takes as input $1^{\ell}$ and outputs the public parameters $ps$.
	\item{\sf KeyGen} $(1^{\ell})\rightarrow (SK,PK).$ The key generation algorithm takes as input $1^{\ell}$ and outputs a secret-public key pair $(SK,PK)$.
	\item{\sf Sign} $(ps,SK,M)\rightarrow \sigma.$ The  signature algorithm takes as input the public parameters $ps$,  the secret ky $SK$ and a message $M$, and outputs a signature $\sigma$ on $M$.
	\item{\sf Verify} $(ps,M,PK,\sigma)\rightarrow True/ False.$  The verification algorithm takes as input the public parameters $ps$, the message $M$, the public key $PK$ and the signature $\sigma$, and outputs $True$ if {\sf Sign} $(ps,M,SK)\rightarrow \sigma$; otherwise, it outputs $False$.
\end{itemize}
\begin{definition}{\sf Correctness.}
	We say that a digital signature is correct if
	\begin{eqnarray*}
		\Pr\left[\begin{array}{l|l}
			& {\sf Setup}(1^{\ell})\rightarrow ps;\\
			{\sf V}(ps,M,PK,\sigma)\rightarrow T & {\sf KeyGen}(1^{\ell})\rightarrow (SK,PK);\\
			& {\sf Sign}(ps,SK,M)\rightarrow \sigma.\end{array}\right]\geq 1-\epsilon(\ell)
	\end{eqnarray*}
	and
	\begin{eqnarray*}
		\Pr\left[\begin{array}{l|l}
			& {\sf Setup}(1^{\ell})\rightarrow ps;\\
			{\sf V}(ps,M,PK,\sigma)\rightarrow F & {\sf KeyGen}(1^{\ell})\rightarrow (SK,PK);\\
			& {\sf Sign}(ps,SK,M)\rightarrow \sigma.\end{array}\right]< \epsilon(\ell)
	\end{eqnarray*}
	where the probability is taken over the random coins consumed by all algorithms in the scheme.
\end{definition}

\noindent{\em Security Model.} A digital signature scheme should achieve the traditional security called {\em existential unforgeability under adaptive chosen message attacks} (EU-CMA) \cite{gmr:88}. This model is formally defined by the following game executed between a challenger $\cC$ and an adversary $\cA$.
\begin{itemize}
	\item{\sf Setup.} $\cC$ runs {\sf Setup} $(1^{\ell})$  to generate the public parameters $params$ and sends them to $\cA$.
	\item{\sf KeyGen.} $\cC$ runs {\sf KeyGen} $(1^{\ell})$  to generate a secret-public pair $(SK,PK)$ and sends $PK$ to $\cA$.
	\item{\sf Query.} $\cA$ can adaptively query the signature oracle. $\cA$ sends a message $M$ to $\cC$. $\cC$ runs {\sf Sign} $(params,SK,M)$ to generate a signature $\sigma$ on $M$ and responds $\cA$ with $\sigma$. This query can be made multiple times.
	\item{\sf Output.} $\cA$ outputs a message-signature pair $(M^{*},\sigma^{*})$. $\cA$ wins the game if $M^{*}$ has not been used to query the signature oracle  and {\sf Verify} $(params, M^{*}, PK, \sigma^{*})\rightarrow True$.
\end{itemize}

\section {Provenance-based Classification Scheme}

In this section, we describe our provenance-based classification scheme. After a brief overview of preliminaries needed for our scheme, we provide details of our scheme which consists of \emph{setup phase} and \emph{verification phase}. Finally, after presenting the security game of a chosen-word-attack, we give the security proof of our scheme showing that it is semantically secure in the next section. 

\subsection{Algorithms}

Let $\mathbb{G}_1, \mathbb{G}_2$ be two cyclic multiplicative groups with the same order $p$. The size of $ \mathbb{G}_1, \mathbb{G}_2$ is determined by the security parameter. Let $\hat{e}:\mathbb{G}_1 \times \mathbb{G}_1 \rightarrow \mathbb{G}_2$ be a bilinear map with the following properties: \\
\begin{itemize}
	\item Bilinearity: $\hat{e}(g_1^a, g_2^b)=\hat{e}(g_1, g_2)^{ab}$ for all $\{g_1, g_2\} \in \mathbb{G}_1, \{a, b \} \in \mathbb{Z}_q$.
	\item Non-degeneracy: There exists $g \in \mathbb{G}_1$ such that $\hat{e}(g, g) \neq 1$.
	\item Computability: There exists an efficient algorithm to compute $\hat{e} (g_1, g_2)$ for all $\{g_1, g_2\} \in \mathbb{G}_1$.
\end{itemize}

The construction of the Provenance-based Classification Scheme is based on identity-based encryption~\cite{BonehF01}. We build a non-interactive searchable encryption scheme from the Bilinear map above and hash functions $H_1: \{0,1\}^* \rightarrow  \mathbb{G}_1$ and $H_2:  \mathbb{G}_2 \rightarrow \{0, 1\}^{\log p}$. In particular, $H_2$ is a collision resistant hash function. The functions in scheme work as follows:

\begin{itemize}
	\item {\sf \emph KeyGen1}: Takes a security parameter $1^\lambda$ as input; then the algorithm picks at random an $\alpha \in Z_p^*$ and a generator 
	$g \in  \mathbb{G}_1$, where $p$ is a prime and it is the size of $ \mathbb{G}_1$ and $ \mathbb{G}_2$. 
	It outputs the public key $A_{pub} = [g, h_1=g^\alpha]$ and the private key
	$A_{priv} = \alpha$. \\
	
	\item {\sf \emph KeyGen2}: Takes a security parameter $1^\lambda$ as input; then the algorithm picks at random a $\beta \in Z_p^*$ and a generator 
	$g \in  \mathbb{G}_1$, where $p$ is a prime and it is the size of $ \mathbb{G}_1$ and $ \mathbb{G}_2$. 
	It outputs the public key $B_{pub} = [g, h_2=g^\beta]$ and the private key
	$B_{priv} = \beta$. \\
	
	\item {\sf \emph PBCT} ($A_{pub}$,$B_{priv}$): Generates a Provenance-based Classification Tags (PBCTs) for provenance fragments for the purpose of searching. Then Computes $t =\hat{e} (H_1({\cal P})^\beta,h_1^r)\in  \mathbb{G}_2$ for a random $r\in Z_p^*$ and a provenance fragment ${\cal P}$.
	Output {\sf PBCT} ($A_{pub},\beta$) = [$h_1^\beta, h_2^r, H_2(t)] \equiv [X,Y,Z]$. \\
	
	\item {\sf \emph Trapdoor} ($A_{priv}$): Output $T_{\cal P}$ = $H_1({\cal P'})^\alpha \in  \mathbb{G}_1$, where $\cal P'$ is provenance fragments chosen by the administrator PAP. \\
	
	\item {\sf \emph Test} ($A_{pub},B_{pub},T_{\cal P}, S$):
	Test if $H_2(\hat{e}(T_{\cal P}, Y)) = Z$
	and $\hat{e}(X, g) = \hat{e}(h_1,h_2)$. If both are true, output 1; otherwise 0. The test function using $A_{pub},B_{pub}$ checks if the encrpyted provenance matching $T_{\cal P}$ satisfies the policies; it also verifies if the provenance is generated by authenticated users by checking the short signature.  \\
	
\end{itemize}

\subsection{Schemes}

Our policy-based classification scheme as shown in the figure below has two phases, namely the setup phase and the verification phase. Initially, in the setup phase, both the administrator PAP and users generate their own pair of public and private keys. Then PAP calculates $Trapdoor$ for sets of provenance fragments listed in the access control policies and sends them with policies to PDP which executes the test function. In the verification phase, before users send encrypted files and provenance to the Cloud Server, they calculate $PBCT$ and attach them to the files. After receiving files, the PDP classifies them by running the $Test$ function. Our scheme involves an encrypted provenance search and is constructed using the technique mentioned in \cite{BonehCOP03}.

\begin {itemize}

\item {\sf Setup Phase}: 

\begin{itemize}
	
	\item PAP runs KeyGen1, taking an input security parameter $1^\lambda$; the algorithm picks a random $\alpha \in Z_p^*$ and a generator $g \in  \mathbb{G}_1$. It outputs the public key $A_{pub} = [g, h_1=g^\alpha]$ and the private key $A_{priv} = \alpha$. Then PAP sends the public keys to users and PDP.
	
	\item Users run KeyGen2 taking an input security parameter $1^\lambda$; the algorithm picks a random $\beta \in Z_p^*$ and a generator $g \in  \mathbb{G}_1$. It outputs the public key $B_{pub} = [g, h_2=g^\beta]$ and the private key $B_{priv} = \beta$. Then, users send public keys to PDP.
	
	\item PAP runs {\sf Trapdoor} ($A_{priv}$) to output $T_{\cal P'} = H_1 ({\cal P})^\alpha \in  \mathbb{G}_1$, and then sends the policies with $T_{\cal P}$ to PDP.
	
\end{itemize}

\item {\sf Verification Phase}: 

\begin{itemize}
	
	\item Users run function {\sf PBCT} ($A_{pub}$,$B_{priv}$) to compute tags where t =\\ $\hat{e}$($H_1({\cal P})^\beta$,$h_1^r$)$\in  \mathbb{G}_2$ for a random $r\in Z_p^*$ and a provenance ${\cal P}$. Output {\sf PBCT}($A_{pub},\beta$) = [$h_1^\beta, h_2^r, H_2(t)] \equiv [X,Y,Z]$. Users then attach the tags with encrypted files and provenance.\\
	
	\item When the encrypted files with tags are sent to PDP, PDP checks if the provenance has the specified keywords in the policies, by running the {\sf Test} function. Test if  $\hat{e}(X, g) \stackrel{?}{=}
	\hat{e} (h_1,h_2)$ (1) and $H_2 (\hat{e}(T_{\cal P}, Y)) \stackrel{?}{=}
	Z$(2). If both are true, then output 1; otherwise 0. The result will then be sent to PEP which executes further operations. \\
	
	\end {itemize}
	
\end{itemize}

\begin{figure*}[th]
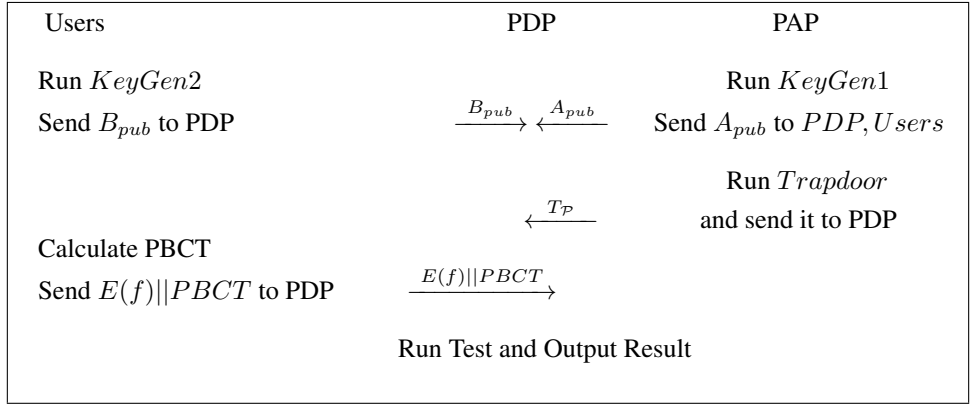

		\begin{center}
			\fbox{
			\begin{tabular}{l@{\hspace{0.5cm}}c@{\hspace{-0.5cm}}l}
				{ Users}    &   { PDP }   &   { ~~~~~~~~~~~~~~~~~PAP }\\
				&    &\\
				Run ${ KeyGen2} $     &    & ~~~~~~~~~~ Run ${KeyGen1}$  \\
				Send $B_{pub}$ to PDP   & $\stackrel{ B_{pub}}{\overrightarrow{\hspace{1cm}}}$ $\stackrel {A_{pub}}{\overleftarrow{\hspace{1cm}}}$  & Send $A_{pub}$ to $PDP, Users$\\
				&& \\
				&&  ~~~~~~~~~~ Run ${Trapdoor}$  \\
				&~~~~~~~~ $\stackrel {T_{\cal P}}{\overleftarrow{\hspace{1cm}}}$ &  ~~~~~~  and send it to PDP\\
				Calculate PBCT &&  \\
				Send $E(f)||PBCT$ to PDP &   $\stackrel{E(f)||PBCT}{\overrightarrow{\hspace{2cm}}}$~~~~~~~~~~~~~~~        & \\
				& &\\
				&~~~  Run Test and Output Result    & \\
				
				& &\\
			\end{tabular}}
		\end{center}
	\caption{Provenance-based Classification Scheme} 
\end{figure*}

In formula (1), the left hand side  $\hat{e} (X, g)= \hat{e} (h_1^\beta,g)= \hat{e} (h_1,g^\beta)$, according to the property of the Bilinear Map. By definition $h_2 = g^\beta$, and hence the left-hand side equals the right-hand side. This formula verifies if the users are authenticated by checking whether they have the corresponding private keys. Similarly, we also prove that the left-hand side equals the right-hand side in formula (2). It tests if PBCT matches the chosen provenance fragments specified by the administrator.\\

\begin{center}
	$H_2(\hat{e}(T_{\cal P}, Y)) \stackrel{?}{=}  Z$ (2)\\
	
	$H_2(\hat{e}(H_1({\cal P})^\alpha,h_2^r)) \stackrel{?}{=} H_2(\hat{e}(H_1({\cal P})^\beta,h_1^r))$\\
	
	$H_2(\hat{e}(H_1({\cal P}),h_2^{\alpha r})) \stackrel{?}{=} H_2(\hat{e}(H_1({\cal P}),h_1^{\beta r}))$\\
	
	$H_2(\hat{e}(H_1({\cal P}),(g^\beta)^{\alpha r})) \stackrel{?}{=}  H_2(\hat{e}(H_1({\cal P}),(g^\alpha)^{\beta r}))$\\
\end{center} 

\section{Complexity Assumptions}

\subsubsection{Discrete Logarithm Assumption}\index{Discrete Logarithm} \index{DL}
The discrete logarithm (DL) assumption \cite{odl:85}  in a finite field is one of the basic assumptions in cryptography research. The DL assumption is defined as follows.
\begin{definition} (Discrete Logarithm (DL) Assumption \cite{odl:85}.)
	Let $\cG(1^{\ell})\rightarrow (p, G)$ and $ G=\langle g\rangle$. Given $(g,y)\in G^{2}$, we say that the discrete logarithm assumption holds on $ G$ if no PPT \index{PPT} adversary $\cA$ can compute a $x\in\bbZp$ such that $y=g^{x}$ with the advantage
	\begin{eqnarray*}
		Adv_{\cA}^{DL}=\Pr\left[y=g^{x}|\cA(p, g,y, G)\rightarrow x\right]\geq \epsilon(\ell)
	\end{eqnarray*}
	where the probability is taken over the random choice of $y\in G$ and the bits consumed by the adversary $\cA$.
\end{definition}

\subsection{Computational Diffie-Hellman Assumption}\index{Computational Diffie-Hellman}\index{CDH}
Diffie and Hellman \cite{dh:76} proposed this assumption and constructed a key exchange scheme based on it. This assumption is defined as follows.
\begin{definition} (Computational Diffie-Hellman (CDH) Assumption \cite{dh:76}.)
	Let $x,y\stackrel{R}{\leftarrow}\bbZp$, $\cG(1^{\ell})\rightarrow (p, G)$ and $ G=\langle g\rangle$. Given $(g,g^{x},g^{y})$, we say that the computational Diffie-Hellman assumption holds on $ G$ if no PPT adversary $\cA$ can compute $g^{xy}$ with the advantage
	\begin{eqnarray*}
		Adv_{\cA}^{CDH}=\Pr\left[\cA( g,g^{x},g^{y})\rightarrow g^{xy}\right]\geq \epsilon(\ell)
	\end{eqnarray*}
	where the probability is taken over the random choices of $x,y\stackrel{R}{\leftarrow}\bbZp$ and the bits consumed by the adversary $\cA$.
\end{definition}

Maurer \cite{mau:94} discussed the relationships between DL assumption and CDH assumption.\index{DL}\index{CDH}

\subsection{Decisional   Diflie-Hellman   Assumption}\index{Decisional Diflie-Hellman}\index{DDH}
Boneh \cite{bon:98} surveyed the various applications of decisional Diffie-Hellman assumption and demonstrated some results regarding it security.
\begin{definition} (Decisional Diflie-Hellman (DDH)   Assumption \cite{bon:98}.)
	Let $x,y,z\stackrel{R}{\leftarrow}\bbZp$, $\cG(1^{\ell})\rightarrow (p, G)$ and $ G=\langle g\rangle$. Given $(g,g^{x},g^{y})$, we say that the decisional Diffie-Hellman assumption holds on $ G$ if no PPT adversary $\cA$ can distinguish $(X,Y,Z)=(g^{x},g^{y},g^{xy})$ from $(X,Y,Z)=(g^{x},g^{y},g^{z})$ with the advantage
	\begin{eqnarray*}
		Adv_{\cA}^{DDH}=\left|\Pr[\cA(X,Y,g^{xy})=  1]-\Pr[\cA(X,Y,g^{z})=1]\right|\geq \epsilon(\ell)
	\end{eqnarray*}
	where the probability is taken over the random choices $x,y,z\stackrel{R}{\leftarrow}\bbZp$ and the bits consumed by the adversary $\cA$.
\end{definition}

\subsection{Computational Bilinear Diffie-Hellman}\index{Computational Bilinear Diffie-\par Hellman}\index{CBDH}
Boneh and Franklin \cite{bf:01} introduced this assumption. This assumption is as follows.
\begin{definition}
	(Computational Bilinear Diffie-Hellman (CBDH) Assumption \cite{bf:01}) Let $\cG\cG(1^{\ell})\rightarrow (e,p, G,  G_T )$ and $ G=\langle g\rangle$. We say that the computational bilinear Diffie-Hellman assumption holds on $(e,p, G,  G_T )$ if no PPT adversaries $\cA$ can compute $e(g,g)^{abc}$ from $(A,B,C)=(g^{a},g^{b},g^{c})$ with the advantage
	\begin{eqnarray*}
		Adv_{\cA}^{CBDH}=\Pr\left[\cA(A,B,C)\rightarrow e(g,g)^{abc}\right]\geq \epsilon(\ell)
	\end{eqnarray*}
	where the probability is taken over the random choices of $a,b,c\stackrel{R}{\leftarrow}\bbZp$ and the bits consumed by $\cA$.
	
\end{definition}

\subsection{Decisional Bilinear Diffie-Hellman Assumption}\index{Decisional Bilinear Diffie-Hellman}\index{DBDH}
Boneh and Franklin \cite{bf:01}  introduced this assumption and used it to construct an identity-based encryption (IBE) scheme. This assumption is defined as follows.
\begin{definition} (Decisional Bilinear Diffie-Hellman (DBDH) Assumption \cite{bf:01})
	Let $a,b,c,z$ $\stackrel{R}{\leftarrow}\bbZp$, $\cG\cG(1^{\ell})\rightarrow(e,p, G,  G_T )$ and  $ G=\langle g\rangle$. We say that the decisional bilinear Diffie-Hellman assumption holds on $(p,e, G,  G_T )$ if no PPT adversary $\cA$ can distinguish $(A,B,C,Z)=(g^{a},g^{b},g^{c},e(g,g)^{abc})$ from $(A,B,C,Z)=(g^{a},g^{b},g^{c},{e(g,g)^{z}})$ with the advantage
	\begin{eqnarray*}
		Adv_{\cA}^{DBDH}=\left|\Pr[\cA(A,B,C,e(g,g)^{abc})=  1]-\Pr[\cA(A,B,C,e(g,g)^{z})=1]\right|\geq \epsilon(\ell)
	\end{eqnarray*}
	where the probability is taken over the random choices of $a,b,c,z\stackrel{R}{\leftarrow}\bbZp$ and the bits consumed by the adversary $\cA$.
	
\end{definition}

\subsection{Symmetric External Diffie-Hellman Assumption}
\index{SXDH}
The Symmetric External Diffie-Hellman (SXDH) assumption \cite{GrothS08}
is defined as follows.

\begin{definition} (Symmetric External Diffie-Hellman Assumption \cite{GrothS08})
	Let $x,y,z$ $\stackrel{R}{\leftarrow}\bbZp$, $\cG\cG(1^{\ell})\rightarrow(e,p, G_1,  G_2, G_T)$
	and  $ G_b=\langle g_b\rangle$ for  any $b \in \{1,2\}$. We say that the Symmetric External Diffie-Hellman Assumption
	holds on $(p,e, G_1, G_2, G_T)$ if no PPT adversary $\cA$ can distinguish $(g_b, g_b^x, g_b^y, g_b^{xy})$ from $(g_b, g_b^x, g_b^y, g_b^z)$ with the advantage
	\begin{eqnarray*}
		Adv_{\cA}^{SXDH}=\left|\Pr[\cA(g_b, g_b^x, g_b^y, g_b^{xy})=1]-  \Pr[\cA(g_b, g_b^x, g_b^y, g_b^z)=1]\right|\geq \epsilon(\ell)
	\end{eqnarray*}
	where the probability is taken over the random choices of $x,y,z\stackrel{R}{\leftarrow}\bbZp$ and the bits consumed by the adversary $\cA$.
	
\end{definition}

\section {Security Proof}

Let us now consider the security proof. This scheme has the property of semantic security against a chosen word attack. That is, PBCT does not reveal any information of provenance to PDP except that ${T_{\cal P}}$ is available to PDP. By simulating our scheme with the game below, an active attacker can obtain  ${T_{\cal P}}$ for any provenance fragment that they choose. However, the attacker could not distinguish PBCT for ${\cal P}_0$ and ${\cal P}_1$ for which it does not know the ${T_{\cal P}}$.\\

\noindent We define the security game between an attacker and the challenger as follows:\\

\noindent \textbf {Provenance-based Classification Security Game:}

\begin{enumerate}
	\item The challenger runs $KeyGen_1$ and $KeyGen_2$ functions to obtain $A_{pub}$, $A_{priv}$ and $B_{pub}$, $B_{priv}$. The challenger then sends  $A_{pub}$ and  $B_{pub}$ to the attacker. 
	\item The attacker sends provenance fragments ${\cal P}$ $\in \{0,1\}^{*}$ of its choice to the challenger. Then the attacker receives trapdoor ${T_{\cal P}}$ calculated by the challenger. 
	\item Then, the attacker sends two random provenance fragments ${\cal P}_0$ and ${\cal P}_1$ for which it did not ask previously ${T_{\cal P}}$.
	\item The challenger chooses a random $b\in \{0,1\}$, and returns C= PBCT ($A_{pub}$, $B_{priv}$, ${\cal P}_b$) to the attacker. 
	\item The attacker can continue to retrieve ${T_{\cal P}}$ from the challenger of any random provenance fragment as long as it is neither {\cal ${\cal P}_0$} nor {\cal ${\cal P}_1$}.
	\item Finally, the attacker makes a guess for $b\in \{0,1\}$ and wins if $b'=b$.
\end{enumerate}

\noindent We define the attacker's advantage to break the Provenance-based Classification Scheme as

\centerline{$Adv(s)=|Pr[b=b']- 1/2|$}

\vspace{0.2in}

To complete the security proof, we define an \textbf{External Bilinear Diffie-Hellman Problem}. We use the $Corollary A.3.$ in~\cite{BonehBG05} to get a new hard problem by setting $P=(1,a, b, c, d, ab, bc)$, $Q=(1)$, $f=abcd$.\\

\noindent \textbf {Corollary A.3.} in ~\cite{BonehBG05}.
Let $P, Q \in F_p[X_1,..,X_n]^s$ be two s-tuples of n-variable polynomials over $F_p$ and let $f \in F_p[X_1,...X_n]$. Let d = max$(2dp, d_Q, d_f)$. If $f$ is independent of (P, Q) then any $A$ that has advantage 1/2 in solving the decision (P,Q,f)-Diffie-Hellman Problem in a generic bilinear group G must take time at least $\Omega (\sqrt{p/d}-s)$. \\

\noindent \textbf {External Bilinear Diffie-Hellman Problem (XBDH)} 
Let $\hat{e}:\mathbb{G}_1 \times \mathbb{G}_1 \rightarrow \mathbb{G}_2$ be a bilinear map. For a generator g of $G_1$, the BDH problem is as follows: given \{$g$, $g^a$, $g^b$, $g^c$, $g^d$, $g^{ab}, g^{bc}\} \in G_1$ as input, compute $\hat{e}(g,g)^{abcd}$.\\

\noindent \textbf {Theorem 1}:  The Provenance-based Classification Scheme given above is semantically secure against a chosen-word-attack in the random oracle model if the XBDH problem is hard.\\

\noindent \textbf {Proof:} Suppose the attacker makes at most $q_1$ hash function queries to $H_2$ and at most $q_2$ trapdoor queries. Assume the attack algorithm has an advantage of $\epsilon$ in breaking the scheme. Then the challenger is able to solve the XBDH problem with an advantage $\epsilon'=\epsilon/ (e{q_1}{q_2})$. We know, in $G_1$, XBDH is a hard problem and $\epsilon'$ is negligible. Therefore, $\epsilon$ must be negligible and the Provenance-based Classification protocol is semantic-secure.

We simulate the game between the attacker $A$ and the challenger $B$. The challenger is given $g$, $u_1 = g^\alpha$, $u_2 = g^\beta$, $u_3 = g^\gamma$, $u_4 = g^\delta$, $u_5 = g^{\alpha\beta}$, $u_6 = g^{\beta\gamma}$. The goal of the challenger is to successfully output $v = e (g,g)^{\alpha\beta\gamma\delta}$. The attacker wins the game if it is able to distinguish $PBCT ({\cal P}_0)$ and $PBCT ({\cal P}_1)$.\\

\noindent {\sf KeyGen}: Challenger sends public keys [$g, u_1, u_2$] to attacker $A$. \\

\noindent{\sf $H_1$-queries}: At anytime, attacker $A$ could query the random oracles $H_1$ by sending a random ${\cal P}_i$, which is a provenance fragment in an item of provenance, while the challenger $B$ keeps a $H_i$-list recorded as $<{\cal P}_i$, $h_i$, $a_j$, $c_j>$ to answer the queries. The list is initially empty. When the attacker $A$ sends ${\cal P}_i \in \{0,1\}^*$ as a query, the challenger $B$ calculates the following:

\noindent 1. The challenger sends $h_i$ back directly as $H_i ({\cal P}_i) = h_i \in G_1$ if ${\cal P}_i$ exists in the current list.\\
2. Or else, the challenger $B$ chooses a random $c_i \in \{0,1\}$, with $Pr[c_i = 0] = 1/ (q_2+1)$.\\
3. Then, the challenger $B$ generates a random $a_i \in Z_p$, and then computes $h_i \leftarrow u_4*g^{a_i} \in G_1$ if $c_i$=0, and  $h_i \leftarrow g^{a_i} \in G_1$ if $c_i$=1.\\
4. Then the challenger adds the newly  generated $<{\cal P}_i$, $h_i$, $a_j$, $c_j>$ to the $H_i$-list and responds to the attacker $h_i$.\\

\noindent{\sf $H_2$-queries}: The attacker $A$ sends t as a $H_2$ query, and the challenger picks a random $V \in \{0,1\}^{log_p}$ as $H_2(t)=V$. Adds the set (t, V) to the $H_2$ list if this does not exist in the list previously. $H_2$ is initially empty.\\

\noindent{\sf Trapdoor queries}: The attacker $A$ sends random ${\cal P}_i$ as trapdoor queries. Then the challenger calculates the following:\\
1. Run $H_1$ query algorithm to obtain $c_i$. If $c_i$=0, it outputs failure and terminates.\\
2. If $c_i$ = 1, outputs $T_i = u_1^{a_i}$ as the result. Note that $T_i = H({\cal P}_i)^\alpha$ as $h_i = g^\alpha$. Then $T_i = H({\cal P}_i)^\alpha$ = ${g^{a_i}}^\alpha$ = $u_1^{a_i}$.\\

\noindent{\sf Challenge}: The attacker $A$ picks two provenance fragments ${\cal P}_0$ and ${\cal P}_1$ for the challenge. Note that both ${\cal P}_0$ and ${\cal P}_1$ should not have been challenged previously. The challenger $B$ calculates PBCT as follows:\\
\noindent 1. The challenger $B$ runs $H_1$-query algorithm to generate $c_0$ and $c_1$: if $c_0$=1 and $c_1$=1, reports failure and terminates; if there is one between $c_0$ and $c_1$ equals 0, then it sets that one to $c_b$; if both of them equal 0, then it randomly chooses one of them to be $c_b$.\\
2. Then it generates a challenge $C$ for ${\cal P}_b$ as $C = [u_5, u_6, J]$, where $J \in \{0,1\}^{log_p}$ is a random number. The challenger $B$ defines $J = H_2(\hat{e}(H_1({\cal P}_b)^\beta, u_1^\gamma))$ = $H_2(\hat{e}(u_4g^{a_b},g^{\alpha\beta\gamma}))$ = 
$H_2(\hat{e}(g,g)^{\alpha\beta\gamma (\delta+a_b)})$\\

\noindent{\sf More trapdoor queries}: The attacker could continue to ask trapdoor for ${\cal P}_i$, where ${\cal P}_i \neq {\cal P}_0$, ${\cal P}_1$. \\

\noindent{\sf Output}: Finally, the attacker $A$ outputs a guess $b' \in \{0,1\}$ which represents whether the challenge $C$ is calculated for ${\cal P}_0$ or ${\cal P}_1$. Then, the challenger chooses a random pair $(t,V)$ from $H_2$ list and calculates $t/\hat{e}(u_5, u_3)^{a_b}$ as the output for $\hat{e}(g,g)^{\alpha\beta\gamma\delta}$, where $a_b$ is known as a parameter to calculate the challenge $C$. \\ 

For the simulation process described above, the probability that a challenger $B$ correctly outputs $\hat{e}(g,g)^{\alpha\beta\gamma\delta}$ is $\epsilon'$. The challenger $B$ wins the game if s/he chooses the correct $H_2$ pair, and does not abort during the trapdoor queries period and the challenge period. \\

\noindent ${\sf Claim1:}$ The probability that a challenger outputs $e(g,g)^{\alpha\beta\gamma\delta}$ is $\epsilon' = \epsilon/ (e{q_1}{q_2})$.\\

\noindent $Proof$: Briefly, the challenger's algorithm does not abort means that it does not abort during either the trapdoor queries period or during the challenge period. The probability that a trapdoor query causes a challenger to abort is $1/(q_2+1)$. Because the attacker makes at most $q_2$ trapdoor queries, the probability that the challenger does not abort at the trapdoor queries phase is at least $(1-1/(q_2 + 1))^{q_T} \geq 1/e$. Similarly, it will abort at the challenge phase when $c_0=c_1=1$ with $Pr[c_0=c_1=1]= (1-1/(q_2+1))^2 \leq 1- 1/{q_2}$. In the opposite way, it does not abort at the challenge phase is at least $1/{q_2}$. Therefore, we have the corresponding probabilities are $Pr[ \xi_1] \geq 1/e$ and $Pr[\xi_2] \geq 1/{q_2}$ respectively. Note that these two events are independent; therefore, the probability that the challenger's algorithm does not abort is $Pr[\xi_1\wedge \xi_2] \geq 1/(e{q_2})$. Following that, the attacker $A$ issues a query for $H_2(e(H_1(W_b)^\beta,u_1^\gamma))$ with probability of at least $\epsilon$; then the challenger chooses the right pair with probability 1/$q_1$. As these processes are independent from one other, we can conclude that the probability that the challenger outputs $\hat{e}(g,g)^{\alpha\beta\gamma\delta}$ is $\epsilon/e{q_1}{q_2}$. As this is a hard problem, the probability of the attacker can be able to break the game is negligible. In other words, the attacker $A$ cannot distinguish whether ${\cal P}_0$ or ${\cal P}_1$ is $PBCT({\cal P}_b)$, hence the Provenance-based Classification Scheme is semantic-secure.

\section{Conclusion}\label{sec:concl}
In this paper, we have proposed a framework which can be used to classify encrypted files sent to the cloud. The classification is made according to the provenance attached to the encrypted files. The provenance information itself is in an encrypted form. The cloud server is able to check whether the provenance satisfies certain policies specified by the administrator without decrypting the provenance. That is, the scheme allows searching encrypted provenance. Furthermore, the cloud server is also able to check the identity of users who sent these files as that is part of the provenance information. We have described the scheme in detail and developed a provenance-based classification security game and proof to show that the proposed scheme is semantically secure based on a hard problem.

However, provenance-based access control is still at its initial stage. There is still interesting work to be done. That will include the examination of the granularity of access control and the range of policy types that can be provided using provenance. By employing provenance, access control systems might support more types of policies beyond the traditional scope. In the meantime, uncertainties might arouse in the evaluation of provenance-based access control policies, especially for a fine-grained approach. Then, conflict solutions might be required.

We also recognise that long-lived and much-handled data can acquire extensive provenance information. In practice, system administrators may need to limit the lifespan of provenance data if this is found to cause unacceptable performance issues. Moreover, a scheme with adaptive semantic-security will improve the security level of the system.

\section{Conclusion}\label{sec:concl}
In this paper, we revisited a privacy-preserving third party
auditing (TPA) cloud storage integrity checking protocol and its
extended version for zero knowledge public auditing (ZKPA). We
showed several security weaknesses in these protocols. It is still an open problem to design a ZKPA protocol that can prevent offline guessing attacks, and we leave it as our future work.

%
%
%
%
\bibliographystyle{my}
\bibliography{reference}
\end{document}